# A Method for Solving Cyclic Block Penta-diagonal Systems of Linear Equations

Milan Batista

University of Ljubljana, Faculty of Maritime Studies and Transport

Pot pomorščakov 4, 6320 Portorož, Slovenia, EU

milan.batista@fpp.edu

**Abstract**

A method for solving cyclic block three-diagonal systems of equations is generalized for solving a block cyclic penta-diagonal system of equations. Introducing a special form of two new variables the original system is split into three block pentagonal systems, which can be solved by the known methods. As such method belongs to class of direct methods without pivoting. Implementation of the algorithm is discussed in some details and the numerical examples are present. The Maple and the Matlab programs are also present in appendices.

**1 Introduction**

The cyclic penta-diagonal systems (CPDS) and cyclic block penta-diagonal systems (CPDS) of linear equations are typically found in numerical solution of one or multidimensional boundary value problems subject to periodic boundary solutions, an approximation of multidimensional periodic functions using splines, etc. These systems can be classified as sparse linear systems ([7]). There exist a relatively large number of good general and/or special purpose programs which can be used for solving these systems via direct or iterative methods ([2],[3],[4],[5],[6],[7],[8]).

In this paper the method that generalizes the method for solving cyclic block three-diagonal system of linear equations ([1],) will be generalized for solving the CBPDS. The algorithm, a description of the implementation, and a numerical example will be presented.

**2 The algorithm**

Consider the CBPDS of linear algebraic equations

$$Ax = f \qquad (1)$$

where

$$A = \begin{bmatrix} C_1 & D_1 & E_1 & 0 & \cdots & A_1 & B_1 \\ B_2 & C_2 & D_2 & E_2 & \ddots & \cdots & A_2 \\ A_3 & B_3 & C_3 & D_3 & E_3 & \ddots & 0 \\ 0 & \ddots & \ddots & \ddots & \ddots & \ddots & \vdots \\ \vdots & \ddots & A_{n-2} & B_{n-2} & C_{n-2} & D_{n-2} & E_{n-2} \\ E_{n-1} & \cdots & \ddots & A_{n-1} & B_{n-1} & C_{n-1} & D_{n-1} \\ D_n & E_n & 0 & \cdots & A_n & B_n & C_n \end{bmatrix} \qquad (2)$$

is a cyclic block penta-diagonal system matrix, $x = (x_1, x_2, ..., x_n)^T$ and $f = (f_1, f_2, ..., f_n)^T$ are unknown and known vectors (RHS vector) respectively and $n \geq 4$ is the number of equations. $A_k$, $B_k$, $C_k$, $D_k$ and $E_k$ are matrices of size $m \times m$, and $f_k$ and $x_k$ are vectors of size $m$, respectively. In what follows, if not stated otherwise, all other matrices have size $m \times m$ and all other vectors have size $m$. Also the vectors and matrices will be denoted by Roman lower-case and upper case letters respectively, and scalars will be denoted by Greek lower-case letters. The unit matrix will be denoted as *I*.

By generalization the idea for the solution of a cyclic block tridiagonal system present in [1], which generalized the method present in [7], we introduce the two unknown vectors $u$ and $v$ of the form

$$u = \alpha(A_1 x_{n-1} + B_1 x_n) + \beta(D_n x_1 + E_n x_2) \qquad v = \gamma A_2 x_n + \delta E_{n-1} x_1 \qquad (3)$$

where are $\alpha$, $\beta$, $\gamma$ and $\delta$ are scalar auxiliary parameters. By inserting (3) into the first two and the last two equations of (1) one obtains the following block penta-diagonal system

$$\tilde{A} x = \tilde{f} \qquad (4)$$

where

$$\tilde{A} = \begin{bmatrix} \tilde{C}_1 & \tilde{D}_1 & E_1 & 0 & \cdots & 0 & 0 \\ \tilde{B}_2 & C_2 & D_2 & E_2 & \ddots & \cdots & 0 \\ A_3 & B_3 & C_3 & D_3 & E_3 & \ddots & 0 \\ 0 & \ddots & \ddots & \ddots & \ddots & \ddots & \vdots \\ \vdots & \ddots & A_{n-2} & B_{n-2} & C_{n-2} & D_{n-2} & E_{n-2} \\ 0 & \cdots & \ddots & A_{n-1} & B_{n-1} & C_{n-1} & \tilde{D}_{n-1} \\ 0 & 0 & 0 & \cdots & A_n & \tilde{B}_n & \tilde{C}_n \end{bmatrix} \qquad (5)$$

$$\tilde{C}_1 = C_1 - \beta/\alpha D_n \qquad \tilde{D}_1 = D_1 - \beta/\alpha E_n \qquad \tilde{B}_2 = B_2 - \delta/\gamma E_{n-1}$$

$$\tilde{D}_{n-1} = D_{n-1} - \gamma/\delta A_2 \qquad \tilde{B}_n = B_n - \alpha/\beta A_1 \qquad \tilde{C}_n = C_n - \alpha/\beta B_1 \qquad (6)$$

$$\tilde{f} = (f_1 - u/\alpha, f_2 - v/\gamma, f_3, \ldots, f_{n-2}, f_{n-1} - v/\delta, f_n - u/\beta)^T$$

By introducing the matrices where matrices $\hat{F} = (I/\alpha, 0, \ldots, 0, I/\beta)^T$ and $\breve{F} = (0, I/\gamma, 0, \ldots, 0, I/\delta, 0)^T$ system (4) can also be written as

$$\tilde{A} x = f - \hat{F} u - \breve{F} v \qquad (7)$$

Inspection of this system suggest that its solution can be sought in the form

$$x = y - U u - V v \tag{8}$$

where $y = (y_1, y_2, ..., y_n)^T$ is new unknown vector and $U = (U_1, U_2, ..., U_n)^T$ and $V = (V_1, V_2, ..., V_n)^T$ are new unknown matrices. Here $y_k$ are vectors, and $U_k$ and $V_k$ are matrices. Substituting (8) into (7) yields

$$(f - \tilde{A}y) + (\tilde{A}U - \hat{F})u + (\tilde{A}V - \breve{F})v = 0 \tag{9}$$

This can be satisfied identically if each term is set equal to zero. In this way one obtains three block penta-diagonal systems with equal system matrices:

$$\tilde{A}y = f \qquad \tilde{A}U = \hat{F} \qquad \tilde{A}V = \breve{F} \tag{10}$$

These systems can be - similarly as block three-diagonal system - solved in three steps ([3],[8]):

- factorization

$$F_1 = C_1^{-1} \qquad P_1 = F_1 D_1 \qquad Q_1 = F_1 E_1$$

$$F_2 = (C_2 - B_2 P_1)^{-1} \qquad P_2 = F_2(D_2 - B_2 Q_1) \qquad Q_2 = F_2 E_2$$

$$H_k = B_k - A_k P_{k-2} \qquad F_k = (C_k - H_k P_{k-1} - A_k Q_{k-2})^{-1} \quad (k = 3,...,n) \tag{11}$$

$$P_k = F_k(D_k - H_k Q_{k-1}) \quad (k = 3,...,n-1)$$

$$Q_k = F_k E_k \quad (k = 3,...,n-2)$$

- intermediate solution

$$g_1 = F_1 f_1 \qquad g_2 = F_2 (f_2 - B_2 g_1)$$

$$g_k = F_k (f_k - H_k g_{k-1} - A_k g_{k-2}) \quad (k = 3,...,n) \tag{12}$$

$$W_1 = F_1/\alpha \qquad W_2 = -F_2 B_2 W_1$$

$$W_k = -F_k (H_k W_{k-1} + A_k W_{k-2}) \quad (k = 3,...,n-1) \tag{13}$$

$$W_n = F_n (I/\beta - H_n W_{n-1} + A_n W_{n-2})$$

$$Z_1 = 0 \qquad Z_2 = F_2/\gamma \qquad Z_3 = -F_3 H_3 Z_2$$

$$Z_k = -F_k (H_k Z_{k-1} + A_k Z_{k-2}) \quad (k = 4,...,n-2,n) \tag{14}$$

$$Z_{n-1} = F_{n-1} (I/\delta - H_{n-1} Z_{n-1} - A_{n-1} Z_{n-3})$$

- Back substitution

$$y_n = g_n \qquad y_{n-1} = g_{n-1} - P_{n-1} y_n$$

$$y_k = g_k - P_k y_{k+1} - Q_k y_{k+2} \quad (k = n-2,...,1) \tag{15}$$

$$U_n = W_n \qquad U_{n-1} = U_{n-1} - P_{n-1} U_n$$

$$U_k = U_k - P_k U_{k+1} - Q_k U_{k+2} \quad (k = n-2,...,1) \tag{16}$$

$$V_n = Z_n \qquad V_{n-1} = V_{n-1} - P_{n-1} V_n$$

$$V_k = V_k - P_k V_{k+1} - Q_k V_{k+2} \quad (k = n-2,...,1) \tag{17}$$

Once systems (10) are solved the equations for computing the unknowns $u$ and $v$ are obtained by substituting (8) into (3) which results in the following auxiliary system of equations

$$\left[I + \alpha\left(A_1 U_{n-1} + B_1 U_n\right) + \beta\left(D_n U_1 + E_n U_2\right)\right] u$$
$$+ \left[\alpha\left(A_1 V_{n-1} + B_1 V_n\right) + \beta\left(D_n V_1 + E_n V_2\right)\right] v = \alpha\left(A_1 y_{n-1} + B_1 y_n\right) + \beta\left(D_n y_1 + E_n y_2\right) \quad (18)$$

$$\left(\gamma A_2 U_n + \delta E_{n-1} U_1\right) u + \left(I + \delta E_{n-1} V_1 + \gamma A_2 V_n\right) v = \gamma A_2 y_n + \delta E_{n-1} y_1$$

After this system is solved the final solution of the system (1) is obtained by (8). Obviously the present algorithm works if all matrices that have to be inverted are non-singular. Also, it is easily established that the proposed algorithm requires approximately $7nm^3 + O(nm^2)$ flops in the factorization step and a total of approximately $36nm^3 + O(nm^2)$ flops.

Before proceeding with the implementation details some remarks regarding choosing values for the parameters $\alpha$, $\beta$, $\gamma$ and $\delta$ will be made. Their values are set in advance so that the non-singularity of $\tilde{C}_1$ and $\tilde{C}_n$ can also be tested in advance; but this is not the case with the solution of system (18) since it includes $U_1, U_2, U_{n-1}, U_n$ and $V_1, V_2, V_{n-1}, V_n$ which are the results of a solution process. If the algorithm is implemented in a symbolic language such as Maple (see Appendix A) then the choice of parameters does not affect the final solution. However if numerical values are used, a bad choice of parameters can make otherwise solvable systems unsolvable; nonetheless, if the values are appropriate they can not affect the final solution if the arithmetic has been performed exactly.

To illustrate the above consideration let take the system of order 5 with the following matrices

$$A_k = \begin{bmatrix} 1 & 1 \\ 1 & -1 \end{bmatrix} \quad B_k = \begin{bmatrix} -1 & 1 \\ 1 & 1 \end{bmatrix} \quad C_k = \begin{bmatrix} 1 & 5 \\ 5 & 1 \end{bmatrix}$$

$$D_k = \begin{bmatrix} 1 & -1 \\ 1 & 1 \end{bmatrix} \quad E_k = \begin{bmatrix} 1 & 1 \\ -1 & 1 \end{bmatrix} \quad f_k = \begin{bmatrix} 10 \\ 10 \end{bmatrix} \quad (k=1,..,5)$$

(19)

This system has the solution $x_k = \begin{bmatrix} 1 & 1 \end{bmatrix}^\mathrm{T}$ $(k=1,..,5)$. It can be shown that in this case

$$u = 2\alpha(1+\lambda)(1,1)^\mathrm{T} \quad v = 2\gamma(1+\sigma)(1,0)^\mathrm{T} \tag{20}$$

where $\lambda \equiv \dfrac{\beta}{\alpha}$ and $\sigma \equiv \dfrac{\delta}{\gamma} \neq 0$ and that the system can be solved with the present algorithm if

$$\lambda \notin \left\{ 0, -3, 4, \frac{13\sigma - 1 \pm \sqrt{2665 + 140\sigma + 40\sigma^2 - \sigma^4}}{\sigma^2 - \sigma - 18}, \right.$$

$$-\frac{12 - 312\sigma + 5\sigma^2 \pm \sqrt{1659 + 15552\sigma + 17016\sigma^2 + 1584\sigma^3 - 839\sigma^4}}{24(\sigma^2 - \sigma - 18)},$$

$$\left. -\frac{522 - 1962\sigma + 33\sigma^2 \pm \sqrt{71085156 - 2470152\sigma + 748344\sigma^2 + 16092\sigma^3 - 30879\sigma^4}}{2(74\sigma^2 - 41\sigma - 1424)} \right\}$$

**3 Implementation**

The implementation of the algorithm can be done in practice with several simplifications regarding computer memory. First, in the factorization step, $B_k$ can be overwritten by $H_k$, $C_k$ can be overwritten by $F_k$, $D_k$ can be overwritten by $P_k$ and $E_k$ can be overwritten by $Q_k$. Note that in this way $B_1$, $D_n$, $E_{n-1}$ and $E_n$ which are needed in the solution of auxiliary system (18) remain unchanged.

Now if the system has several RHSs, then extra memory of approximately $m^2 n$ is needed for storing matrices $W_k$ and $Z_k$, which can then be overwritten in the back-substitution phase by matrices $U_k$ and $V_k$. In this case one should, in the factorization step, also solve the second and the third system of equations (10) - i.e., complete steps (13), (14), and then (16) and (17) - which are independent particular rhs.of the system. Note that in the factorization phase one should also compute the inversion of the auxiliary system matrices (18). Consequently no matrices inversion is needed in the solution phase. Thus for particular a RHS one should only perform the intermediate and back substitution phase for the system (10)--i.e., steps (12) and (15)--then complete the solution of (18) and finally obtain the solution by (8).

If one must, however, solve the system with only one RHS then in the intermediate, back substitution and final solution steps $f_k$ can be overwritten by $g_k$, then by $y_k$ and at last with $x_k$, $C_k$ can be overwritten by $W_k$ and then with $U_k$, and $B_k$ can be overwritten first by $Z_k$ and then by $V_k$. Note that in this case $B_1$ should first be stored in temporary memory. Thus if the factorization, intermediate solution and final solution steps of the algorithm are not separated, then no extra work storage is necessary in addition to the memory required for the system matrix and rhs vector, except the storage for the solution of the auxiliary system (18) and storage for a copy of $B_1$. The examples of a practical implementation in Maple and in Matlab with the above simplifications are provided in the Appendix A and B . The programs are freely available at. http://www.fpp.edu/~milanb/penta

**4 Numerical examples**

In this section four simple numerical examples will be presented which will illustrate the performance of the algorithm. For numerical computation the algorithm was implemented in Fortran 90 and MATLAB using double precision arithmetic. All the computations were executed on a PC with an Intel Pentium 4 CPU 3.4GHz 2GB RAM processor. In examples errors and residuals are computed respectively as $Err = \|x - \tilde{x}\|_\infty$ and $Res = \|f - A\tilde{x}\|_\infty$ where $\tilde{x}$ is the computed solution, and the execution time in seconds was recorded. If not otherwise stated all the tested examples assume the exact solution $x_k = 1$.

**Example 1.** As first example, the numerical performance on random generated matrices is tested. The standard Fortran 90 intrinsic subroutine *random number* was used for the random generation of $A_k$, $B_k$, $C_k$, $D_k$ and $E_k$ and the value of $\sigma = 4m$ was added to diagonals of $C_k$. The results are presented in Table 1. Here the accuracy of the solution is pure. The error drops to 2.18e-05 if $\sigma = 100m$ and only to 2.20e-07 if $\sigma = 1000m$. However in all the cases the average error is below $10^{-5}$.

**Table 1.** Execution time, errors and residuals for random matrices for $n = 100\,000$ and $\alpha = \gamma = 1$, $\beta = \delta = -1$ and $\sigma = 4m$

| m | Err | Res | Time |
|---|---|---|---|
| 2 | 1.1802e-2 | 9.9615e-02 | 0.156 |
| 4 | 0.86270e-3 | 0.1360 | 0.625 |
| 8 | 2.2462e-3 | 7.1822e-02 | 3.219 |

**Example 2.** For comparison, consider the example of Hermitian CBPS from [5] where $A_k = E_k = I$, $B_k = D_k$, and

$$C_k = circ(22, -8, 1, ..., 1, -8) \qquad B_k = circ(-7.2, 1.8, ...., 1.8) \qquad (21)$$

are circulant matrices. For calculations both MATLAB and Fortran implementation was used so the execution time can be compared. The results are given in Table 2. Comparing with results from [5] (Table 2) one can see that the MATLAB execution time of algorithms are almost similar however the special purposed algorithms in [5] give more accurate results with error of order $10^{-14}$ while the error of present algorithm is of order $10^{-10}$.

14.3.2008  8:32:55                              9                              arXiv:0803.0874

**Table 2.** The error, residual and execution time for Example 1 with $m = 7$, $\alpha = \gamma = 1$, and $\beta = \delta = -1$.

|  | Matlab | | | Fortran90 | | |
| --- | --- | --- | --- | --- | --- | --- |
| n | Err | Res | Time | Err | Res | Time |
| 500 | 3.5083e-014 | 5.4357e-013 | 0.157 | 3.0931e-13 | 4.6896e-12 | 0.000 |
| 1000 | 6.7724e-014 | 6.0041e-013 | 0.265 | 3.3129e-13 | 4.6895e-12 | 0.031 |
| 2000 | 1.3350e-010 | 9.4539e-010 | 0.563 | 7.4600e-11 | 3.5513e-10 | 0.047 |
| 4000 | 1.5404e-010 | 9.1974e-010 | 1.766 | 4.4544e-11 | 3.5117e-10 | 0.078 |
| 6000 | 6.5362e-011 | 4.7643e-010 | 3.266 | 3.4862e-11 | 3.4581e-10 | 0.125 |
| 8000 | 5.6390e-011 | 4.7620e-010 | 5.219 | 3.5578e-11 | 3.5013e-011 | 0.172 |
| 16000 | - | - | - | 6.4291e-10 | 3.6672e-10 | 0.344 |
| 32000 | - | - | - | 4.0685e-10 | 5.1651e-10 | 0.688 |
| 64000 | - | - | - | 8.5715e-10 | 8.4607e-10 | 1.359 |

**Example 3.** As a last example, consider the following simple two-point boundary value problem: find a function $y_1(x)$ and $y_2(x)$, $x \in [0,1]$ satisfying the system of ordinary differential equations

$$y_1'' + y_2 = \cos 2\pi x - 4\pi^2 \sin 2\pi x$$
$$y_2'' + y_1 = \sin 2\pi x - 4\pi^2 \cos 2\pi x \quad (22)$$

with periodic boundary condition

$$y_1(0) = y_1(1) \quad y_1'(0) = y_1'(1) \quad y_2(0) = y_2(1) \quad y_2'(0) = y_2'(1) \quad (23)$$

The exact solution of the system is

$$y_{1,ex} = \sin 2\pi x \quad y_{2,ex} = \cos 2\pi x \quad (24)$$

By dividing the interval $[0,1]$ into $n$ equal intervals of length $h = \dfrac{1}{n}$, and using fourth-order difference approximation for second order derivatives of unknowns

$$y_j''(x_k) = y_{j,k}'' \simeq \frac{-y_{j,k-2} + 16y_{j,k-1} - 30y_{j,k} + 16y_{j,k+1} - y_{j,k+2}}{12h^2} \quad (j = 1,2) \quad (25)$$

one obtains the cyclic block penta-diagonal system for $2n$ unknowns $y_{j,k} = y_j(x_k)$, $j = 1, 2$, $k = 1,..,n$ with

$$A_k = E_k = \begin{bmatrix} -1 & 0 \\ 0 & -1 \end{bmatrix} \quad B_k = D_k = \begin{bmatrix} 16 & 0 \\ 0 & 16 \end{bmatrix} \quad C_k = \begin{bmatrix} -30 & 12h^2 \\ 12h^2 & -30 \end{bmatrix}$$

$$f_k = 12h^2 \begin{bmatrix} \cos(2\pi x_k) - 4\pi^2 \sin(2\pi x_k) \\ \sin(2\pi x_k) - 4\pi^2 \cos(2\pi x_k) \end{bmatrix} \quad x_k = (k-1)h \quad (k = 1,...,n)$$

(26)

The above system is solved with several different numbers of intervals. At each run the $Err$ and average error $\bar{\varepsilon}$ are recorded

$$\bar{\varepsilon} = \frac{1}{n} \sum_{j=1}^{2} \sum_{k=1}^{n} |y_{j,k} - y_{j,ex}|$$

(27)

The results are presented in Table 3. As can be seen from the table, the average error decreases with the number of intervals $n$. A somewhat detailed analysis shows that $\bar{\varepsilon}$ decreases as $11 \times h^4$, which confirms the theoretical results that with finer mesh the error should vanish proportionally to $h^4$

**Table 3.** The error and average error for Example 3
for different number of intervals $n$.

| $n$ | 20 | 40 | 80 | 160 | 320 | 640 |
|---|---|---|---|---|---|---|
| $Err$ | 1.074e-4 | 6.754e-5 | 4.228e-7 | 2.644e-8 | 1.654e-9 | 1.053e-10 |
| $\bar{\varepsilon}$ | 6.806e-5 | 4.299e-6 | 2.693e-7 | 1.684e-8 | 1.052e-9 | 6.581e-11 |

**Conclusion**

The algorithm that extends the method for solving cyclic block tridiagonal systems to cyclic block penta-diagonal systems is introduced. The algorithm requires approximately $36nm^3$ flops and is thus comparable with the other, similar, algorithms. The results of numerical examples show that the presented algorithm produces relatively accurate results within an acceptable execution time. The advantage of the presented algorithm is that it is general, simple and relatively easy to program.

**Acknowledgement.** The author wishes to thank Dr. A.A.Karawia, from Mathematics Department, Faculty of Science, Mansoura University, Egypt, who initiate the present research.

**Appendix A. The Maple procedure for solving cyclic block penta-diagonal linear systems of equations**

```
> restart:
> with(linalg):
Warning, the protected names norm and trace have been redefined and
unprotected
```

```
> #*****************************************
> # CBPS - Cyclic Block Penta-diagonal Solver
> # History: MB Created 7.3.2008
> #*****************************************
> cycpen:=proc( m, n, EE, AA, BB, CC, DD, FF)
> local k, alpha, beta, gamma1, delta, lambda, sigma, K,
BB1, TT, A11, A12, A21, A22, B1, B2, u, v:
> #================================
> # (1) Initialization
> #================================
> K:=matrix(m,m,0):for k from 1 to m do;K[k,k]:=1;od:
> #alpha:=1:beta:=1;gamma1:=1;delta:=1;
> beta:=lambda*alpha:
> delta:=sigma*gamma1:
> BB1:=BB(1):
> CC(1):=evalm(CC(1)-beta/alpha*DD(n)):
> DD(1):=evalm(DD(1)-beta/alpha*EE(n)):
> BB(2):=evalm(BB(2)-delta/gamma1*EE(n-1)):
> DD(n-1):=evalm(DD(n-1)-gamma1/delta*AA(2)):
> BB(n):=evalm(BB(n)-alpha/beta*AA(1)):
> CC(n):=evalm(CC(n)-alpha/beta*BB(1)):
> #================================
> # (2) Factorization
> #================================
> CC(1):=inverse(CC(1)):
> DD(1):=simplify(evalm(CC(1)&*DD(1))):
> EE(1):=simplify(evalm(CC(1)&*EE(1))):
> CC(2):=simplify(inverse(CC(2)-evalm(BB(2)&*DD(1)))):
> DD(2):=simplify(evalm(CC(2)&*evalm(DD(2)-
evalm(BB(2)&*EE(1))))):
> EE(2):=simplify(evalm(CC(2)&*EE(2))):
> for k from 3 to n do; BB(k):=simplify(evalm(BB(k)-
evalm(AA(k)&*DD(k-
2)))):CC(k):=simplify(inverse(evalm(CC(k)-
evalm(BB(k)&*DD(k-1))-evalm(AA(k)&*EE(k-2))))):if (k < n)
then DD(k):=simplify(evalm(CC(k)&*evalm(DD(k)-
evalf(BB(k)&*EE(k-1))))) end if: if (k < n-1) then
EE(k):=simplify(evalm(CC(k)&*EE(k))): end if:  od:
> #================================
> # (3) Intermediate solution
> #================================
> FF(1)  :=simplify(evalm(CC(1)&*FF(1))):
> CC(1):=simplify(evalm(CC(1)/alpha)):
> BB(1):=matrix(m,m,0):
> TT:=CC(2):
> FF(2):=simplify(evalm(TT&*(FF(2)-evalm(BB(2)&*FF(1))))):
> CC(2):=simplify(evalm(-TT&*evalm(BB(2)&*CC(1)))):
> BB(2):=evalm(TT/gamma1):
> for k from 3 to n do: TT:=CC(k):
FF(k):=simplify(evalm(TT&*evalm(FF(k)-evalm(BB(k)&*FF(k-
1))-evalm(AA(k)&*FF(k-2))))):CC(k):=simplify(evalm(-
TT&*(evalm(BB(k)&*CC(k-1))+evalm(AA(k)&*CC(k-
2))))):BB(k):=simplify(evalm(-TT&*(evalm(BB(k)&*BB(k-
1))+evalm(AA(k)&*BB(k-2))))): if (k = n) then
```

```
   CC(k):=simplify(evalm(CC(k)+TT/beta)): end if: if (k = n-1)
   then BB(k):=simplify(evalm(BB(k)+TT/delta)): end if: od:
> #==================
> # (4) Back substitution
> #==================
> FF(n-1):=simplify(evalm(FF(n-1)-evalm(DD(n-1)&*FF(n)))):
> CC(n-1):=simplify(evalm(CC(n-1)-evalm(DD(n-1)&*CC(n)))):
> BB(n-1):=simplify(evalm(BB(n-1)-evalm(DD(n-1)&*BB(n)))):
> for k from n-2 by -1 to 1 do;
k;FF(k):=simplify(evalm(FF(k)-evalm(DD(k)&*FF(k+1))-
evalm(EE(k)&*FF(k+2)))): CC(k):=simplify(evalm(CC(k)-
evalm(DD(k)&*CC(k+1))-
evalm(EE(k)&*CC(k+2))));BB(k):=simplify(evalm(BB(k)-
evalm(DD(k)&*BB(k+1))-evalm(EE(k)&*BB(k+2))));od:
> #============================
> # (5) Solution of aux. system
> #============================
> A11:=simplify(evalm(K+alpha*(evalm(AA(1)&*CC(n-
1))+evalm(BB1&*CC(n)))+beta*(evalm(DD(n)&*CC(1))+evalm(EE(n
)&*CC(2))))):
> A12:=simplify(evalm(alpha*(evalm(AA(1)&*BB(n-
1))+evalm(BB1&*BB(n)))+beta*(evalm(DD(n)&*BB(1))+evalm(EE(n
)&*BB(2))))):
>
A21:=simplify(evalm(gamma1*evalm(AA(2)&*CC(n))+delta*evalm(
EE(n-1)&*CC(1)))):
>
A22:=simplify(evalm(K+gamma1*evalm(AA(2)&*BB(n))+delta*eval
m(EE(n-1)&*BB(1)))):
> B1:=simplify(evalm(alpha*(evalm(AA(1)&*FF(n-
1))+evalm(BB1&*FF(n)))+beta*(evalm(DD(n)&*FF(1))+evalm(EE(n
)&*FF(2))))):
>
B2:=simplify(evalm(gamma1*evalm(AA(2)&*FF(n))+delta*evalm(E
E(n-1)&*FF(1)))):
> A11:=simplify(inverse(A11)):
> v:=simplify(evalm(inverse(A22-A21&*(A11&*A12))&*(B2-
evalm(evalm(A21*A11))&*B1))):
> u:=simplify(evalm(A11&*(B1-evalm(A12&*v)))):
> #=======================================
> # (6) Final solution
> #=======================================
> for k from 1 to n do; FF(k) := simplify(evalm(FF(k) -
evalm(CC(k)&*u)-evalm(BB(k)&*v))): od:
> #=====END BCBP
> end proc:
>
> #********************************
> # Example
> #*********************
> m:=2:n:=5:
> # Form LHS
> EE:=k->matrix(m,m,[1,1,1,-1]):
> AA:=k->matrix(m,m,[-1,1,1,1]):
> BB:=k->matrix(m,m,[1,5,5,1]):
```

```
> CC:=k->matrix(m,m,[1,-1,1,1]):
> DD:=k->matrix(m,m,[1,1,-1,1]):
> FF:=k->vector(m,[]):
> # Form RHS
> for k from 1 to n do; FF(k) :=
evalm((EE(k)+AA(k)+BB(k)+CC(k)+DD(k))&*vector(m,[1,1])):
od:
> # Solve system
> cycpen(m,n,EE,AA,BB,CC,DD,FF):
> for k from 1 to n do; FF(k): od;
                                [1, 1]
                                [1, 1]
                                [1, 1]
                                [1, 1]
                                [1, 1]
>
```

## Appendix B. The Matlab program for solving cyclic block penta-diagonal linear systems of equations

```
function X = CBPS( A, B, C, D, E, F, par)
% Cyclic Block Penta-diagonal Linear System Solver
% History: 20.jan.2008 MB Created
    alp = 1;
    bet = -1;
    gam = 1;
    del = -1;
    if (nargin == 7)
        if (length(par) == 4)
            alp = par(1);
            bet = par(2);
            gam = par(3);
            del = par(4);
        end
    end
% Local arrays
    [m,m1,n]=size(A);
    T = zeros(m,m);
    S = eye(2*m,2*m);
    P = zeros(2*m,1);
    B1(:,:) = B(:,:,1);
% Initialization
    C(:,:,1) = C(:,:,1) - (bet/alp)*D(:,:,n);
    D(:,:,1) = D(:,:,1) - (bet/alp)*E(:,:,n);
    B(:,:,2) = B(:,:,2) - (del/gam)*E(:,:,n-1);
    D(:,:,n-1) = D(:,:,n-1) - (gam/del)*A(:,:,2);
    B(:,:,n) = B(:,:,n) - (alp/bet)*A(:,:,1);
    C(:,:,n) = C(:,:,n) - (alp/bet)*B(:,:,1);
% Factorization
    C(:,:,1) = inv(C(:,:,1));
    D(:,:,1) = C(:,:,1)*D(:,:,1);
    E(:,:,1) = C(:,:,1)*E(:,:,1);
    C(:,:,2) = inv(C(:,:,2) - B(:,:,2)*D(:,:,1));
    D(:,:,2) = C(:,:,2)*(D(:,:,2) - B(:,:,2)*E(:,:,1));
```

```matlab
        E(:,:,2) = C(:,:,2)*E(:,:,2);
        for k =3:n
            B(:,:,k) = B(:,:,k) - A(:,:,k)*D(:,:,k-2);
            C(:,:,k) = inv(C(:,:,k) - B(:,:,k)*D(:,:,k-1) -
A(:,:,k)*E(:,:,k-2));
            if (k < n)
                D(:,:,k) = C(:,:,k)*(D(:,:,k)-B(:,:,k)*E(:,:,k-1));
            end
            if (k < n - 1)
                E(:,:,k) = C(:,:,k)*E(:,:,k);
            end
        end
% Intermediate solution
    F(:,1) = C(:,:,1)*F(:,1);
    C(:,:,1) = C(:,:,1)/alp;
    B(:,:,1) = 0;
    T = C(:,:,2);
    F(:,2)   = T*(F(:,2) - B(:,:,2)*F(:,1));
    C(:,:,2) = -T*(B(:,:,2)*C(:,:,1));
    B(:,:,2) = T/gam;
    for k = 3:n
        T = C(:,:,k);
        F(:,k) = T*(F(:,k)-B(:,:,k)*F(:,k-1)-A(:,:,k)*F(:,k-2));
        C(:,:,k) = -T*(B(:,:,k)*C(:,:,k-1)+A(:,:,k)*C(:,:,k-2));
        B(:,:,k) = -T*(B(:,:,k)*B(:,:,k-1)+A(:,:,k)*B(:,:,k-2));
        if (k == n)
            C(:,:,k) = C(:,:,k)+T/bet;
        end
        if (k == n-1)
            B(:,:,k) = B(:,:,k)+T/del;
        end
    end
% Back substitution
    F(:,n-1) = F(:,n-1)-D(:,:,n-1)*F(:,n);
    C(:,:,n-1) = C(:,:,n-1)-D(:,:,n-1)*C(:,:,n);
    B(:,:,n-1) = B(:,:,n-1)-D(:,:,n-1)*B(:,:,n);
    for k = n-2:-1:1
        F(:,k) = F(:,k) - D(:,:,k)*F(:,k+1) - E(:,:,k)*F(:,k+2);
        C(:,:,k) = C(:,:,k) - D(:,:,k)*C(:,:,k+1) -
E(:,:,k)*C(:,:,k+2);
        B(:,:,k) = B(:,:,k) - D(:,:,k)*B(:,:,k+1) -
E(:,:,k)*B(:,:,k+2);
    end
% Solution of auxiliary system
    m1 = m+1;
    S(1:m,1:m) = S(1:m,1:m) + alp*(A(:,:,1)*C(:,:,n-
1)+B1(:,:)*C(:,:,n))+bet*(D(:,:,n)*C(:,:,1)+E(:,:,n)*C(:,:,2));
    S(1:m,m1:2*m) = alp*(A(:,:,1)*B(:,:,n-
1)+B1(:,:)*B(:,:,n))+bet*(D(:,:,n)*B(:,:,1)+E(:,:,n)*B(:,:,2));
    S(m1:2*m,1:m) = gam*A(:,:,2)*C(:,:,n) + del*E(:,:,n-1)*C(:,:,1);
    S(m1:2*m,m1:2*m) = S(m1:2*m,m1:2*m) + gam*A(:,:,2)*B(:,:,n) +
del*E(:,:,n-1)*B(:,:,1);
    P(1:m) = alp*(A(:,:,1)*F(:,n-
1)+B1(:,:)*F(:,n))+bet*(D(:,:,n)*F(:,1)+E(:,:,n)*F(:,2));
    P(m1:2*m) = gam*A(:,:,2)*F(:,n) + del*E(:,:,n-1)*F(:,1);
    P = S\P;
% Final solution
    for k = 1:n
        X(:,k) = F(:,k) - C(:,:,k)*P(1:m) - B(:,:,k)*P(m1:2*m);
```

```
        end
return
```